\documentclass[a4paper,twocolumn, nofootinbib,aps,prb,eqsecnum,notitlepage,showkeys]{revtex4-2}
\usepackage{multirow}
\usepackage{graphicx}
\usepackage{bm}
\usepackage{array}
\usepackage{comment}
\usepackage{dcolumn}
\usepackage{hyperref}
\usepackage{gensymb}
\usepackage{amsmath}
\usepackage{dcolumn}    
\usepackage{ifpdf}
\usepackage{amssymb,lineno,amsfonts}
\usepackage{graphicx}   
\usepackage{bm}         
\usepackage{bbm}
\usepackage{mathrsfs}
\usepackage{upgreek}
\usepackage{mathtools}
\usepackage{epstopdf}
\usepackage{setspace}
\usepackage{hyperref}
\usepackage{natbib}
\usepackage{multirow}
\usepackage{esvect}
\usepackage{soul}
\usepackage{amsmath}
\usepackage[usenames,dvipsnames]{xcolor}
\definecolor{med-blue}{RGB}{0,78,255}
\hypersetup{colorlinks, linkcolor={blue},citecolor={Blue}, urlcolor={blue}}
\usepackage{times}
\usepackage [english]{babel}
\usepackage [autostyle, english = american]{csquotes}
\MakeOuterQuote{"}
\begin{document}
\title{Tensile and compressive strain tuning of a Kondo lattice}
\author{Soumendra Nath Panja}
\email{soumendra.panja@uni-a.de}
\address{Experimental Physics VI, Center for Electronic Correlations and Magnetism,\\
	University of Augsburg, 86159 Augsburg, Germany}
\author{Anton Jesche}
\affiliation{Experimental Physics VI, Center for Electronic Correlations and Magnetism,\\
	University of Augsburg, 86159 Augsburg, Germany}
\author{Nan Tang}
\affiliation{Experimental Physics VI, Center for Electronic Correlations and Magnetism,\\
	University of Augsburg, 86159 Augsburg, Germany}
\author{Philipp Gegenwart}
\email{philipp.gegenwart@uni-a.de}
\affiliation{Experimental Physics VI, Center for Electronic Correlations and Magnetism,\\
	University of Augsburg, 86159 Augsburg, Germany}
\date{\today}
\begin{abstract} 
We present electrical resistivity measurements on the prototypical heavy-fermion metal YbRh$_{2}$Si$_{2}$ (YRS) under $a$-axis tensile and compressive strain and focus on the evolution of the resistivity maximum near 136~K that arises from the interplay of the Kondo effect and the crystal electric field (CEF) splitting. While compressive strain reduces $T_{\rm max}$, similar as previously reported for hydrostatic pressure, $T_{\rm max}$ is enhanced up to 145~K for 0.13\% tensile strain. Model calculations for the strain effect on CEF splitting in YRS reveal a negligible shift of the levels. Instead, the  enhancement of the resistivity maximum indicates a 20\% increase of the Kondo temperature. This opens the perspective to access the hidden zero-field QCP in pure YRS.
\end{abstract}
\pacs{Pacs}
\maketitle
\section{Introduction} 
Heavy-fermion metals with partially filled 4f or 5f shells (typically in Ce, Pr, Yb or U compounds) are prototype materials for the study of quantum criticality~\cite{vLoehneysen,Gegenwart2008}. The Doniach diagram~\cite{Doniach1977}, shown in Fig. 1, illustrates the competition of the Kondo effect, favoring a paramagnetic heavy-fermion state and the indirect exchange (RKKY) coupling, mediating long-range magnetic order. Respective characteristic energies $T_{\rm K}$ and $T_{\rm RKKY}$ depend exponentially and quadratically on the antiferromagnetic (AFM) exchange coupling $J$ between the f-moments and conduction electrons, respectively. The variation of $T_{\rm N}(J)$ reflects this competition, leading to a quantum critical point (QCP)~\cite{Sachdev2011} at a critical $J_{\rm c}$, separating the AFM and paramagnetic (PM) ground states. Experimentally, $J$ can be directly modified by pressure or chemical substitution. While pressure suppresses AFM order for Ce-systems it acts oppositely for Yb-based HF metals, because it favors the smaller magnetic Yb$^{3+}$ configuration over the mixed valent and non-magnetic Yb$^{2+}$ cases. Motivated by the invention and commercialization of in-situ piezoelectric strain tuning tools for low-temperature experiments~\cite{Hicks2014Rsi,razorbill,Barber2019}, we explore in this paper the possibility to {\it enhance} the Kondo coupling of AFM Yb-based HF metals by tensile strain. 

We focus our attention on the prototype Yb-based HF metal YbRh$_{2}$Si$_{2}$ (YRS) which shows a weak AFM ground state below $T_{\rm N}=70$~mK, providing a platform to study quantum criticality~\cite{int3}. A small critical magnetic field of 0.06~T (0.66~T) perpendicular (parallel) to the tetragonal c-axis is sufficient to suppress the AFM order and induce quantum criticality~\cite{int4}. Measurements of the Hall effect detected an additional crossover scale at $T^\ast(B)$ beyond the critical field that has been interpreted as finite temperature signature of a jump of the Fermi surface at the QCP due to a Kondo breakdown~\cite{Paschen}. However, this crossover signals a polarization of ferromagnetic fluctuations~\cite{Gegenwart2005,Tokiwa2009}, distinct from the AFM QCP at zero field~\cite{Schubert}. Since $T_{\rm N}$ at $B=0$ is enhanced under hydrostatic pressure, an extrapolated {\it negative} critical pressure $P_c \approx -0.3$~GPa~\cite{Mederle2002}, corresponding to a relative volume expansion of $+1.6\cdot 10^{-4}$, is required to access the zero-field AFM QCP. The linear thermal expansion coefficient, which is proportional to the uniaxial pressure derivative of the entropy, is negative both along and perpendicular to the tetragonal c-axis in YRS~\cite{Kuechler2004}. The negative Gr\"uneisen ratio of thermal expansion to specific heat indeed implies that the characteristic temperature (here $T_{\rm K}$) rises with volume expansion~\cite{Gegenwart2016}. Since the in-plane thermal expansion exceeds that along the [001] direction by $\sim 80\%$ at low temperature~\cite{Kuechler2004} we expect a significant increase of $T_{\rm K}$ by tensile in-plane strain. YRS single crystals typically grow as thin plates perpendicular to the c-axis, thus the material seems well suitable for exploration.

Here we report on electrical resistance measurements under tensile and compressive piezo-strain tuning, revealing a significant variation of the Kondo maximum temperature, that can be related to a strain dependence of $T_{\rm K}$. Tensile strain significantly enhances $T_{\rm K}$, paving the way to reach a zero-field QCP.

\begin{figure}
	\includegraphics[scale=0.45]{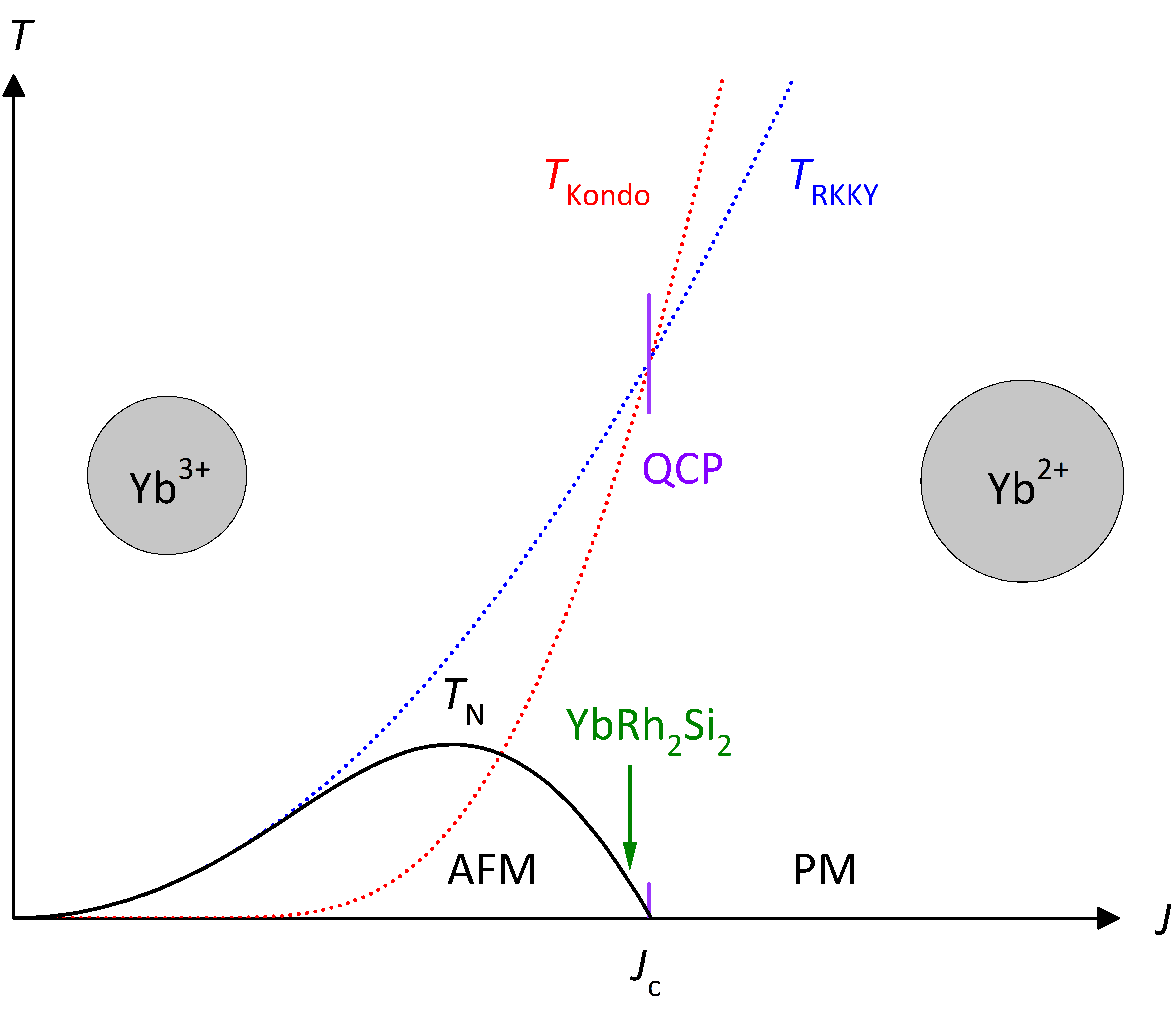}
	\caption{Doniach diagram~\cite{Doniach1977} illustrating $T_{\rm Kondo}$, $T_{\rm RKKY}$, and $T_{\rm N}$ as a function of the exchange coupling $J$ between f- and conduction electrons in heavy-fermion metals. The dashed line at $J_{\rm c}$ indicates the quantum critical point at $T_{\rm N}=0$, separating the antiferromagnetic (AFM) and paramagnetic (PM) ground states. The arrow marks qualitatively the location of YbRh$_2$Si$_2$. The gray spheres illustrate the atomic volume of tri- and divalent Yb.}
	\label{figure1}
\end{figure}


\section{Experimental}

\begin{figure}
	\includegraphics[scale=0.50]{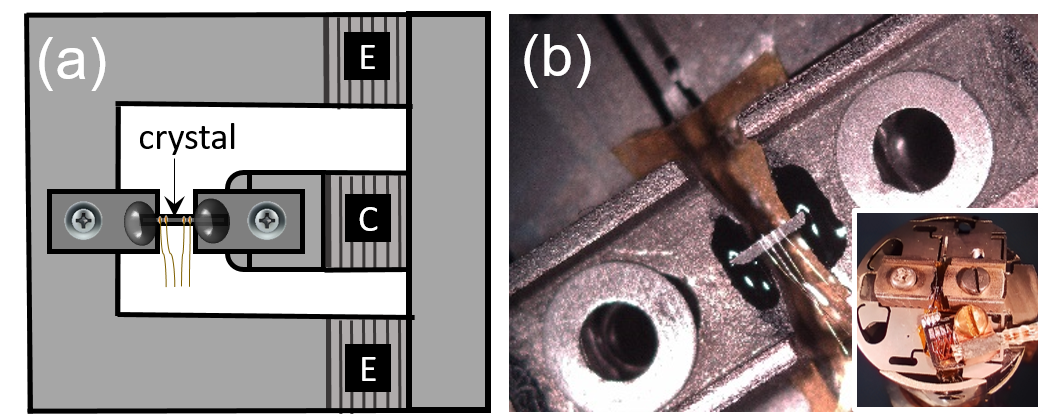}
	\caption{Schematic depiction of the strain apparatus (a). $\bf{E}$ represent the extension stacks and $\bf{C}$ the compression stacks. Panel (b) shows a snapshot taken in the process of sample mounting. Depicting crystal mounted with Stycast on the sample platform of stress cell for resistance measurement. The inset displays the stress cell after sample mounting, ready for the electrical transport measurements.}
	\label{figure2}
\end{figure}

\begin{figure}
	\includegraphics[scale=0.28]{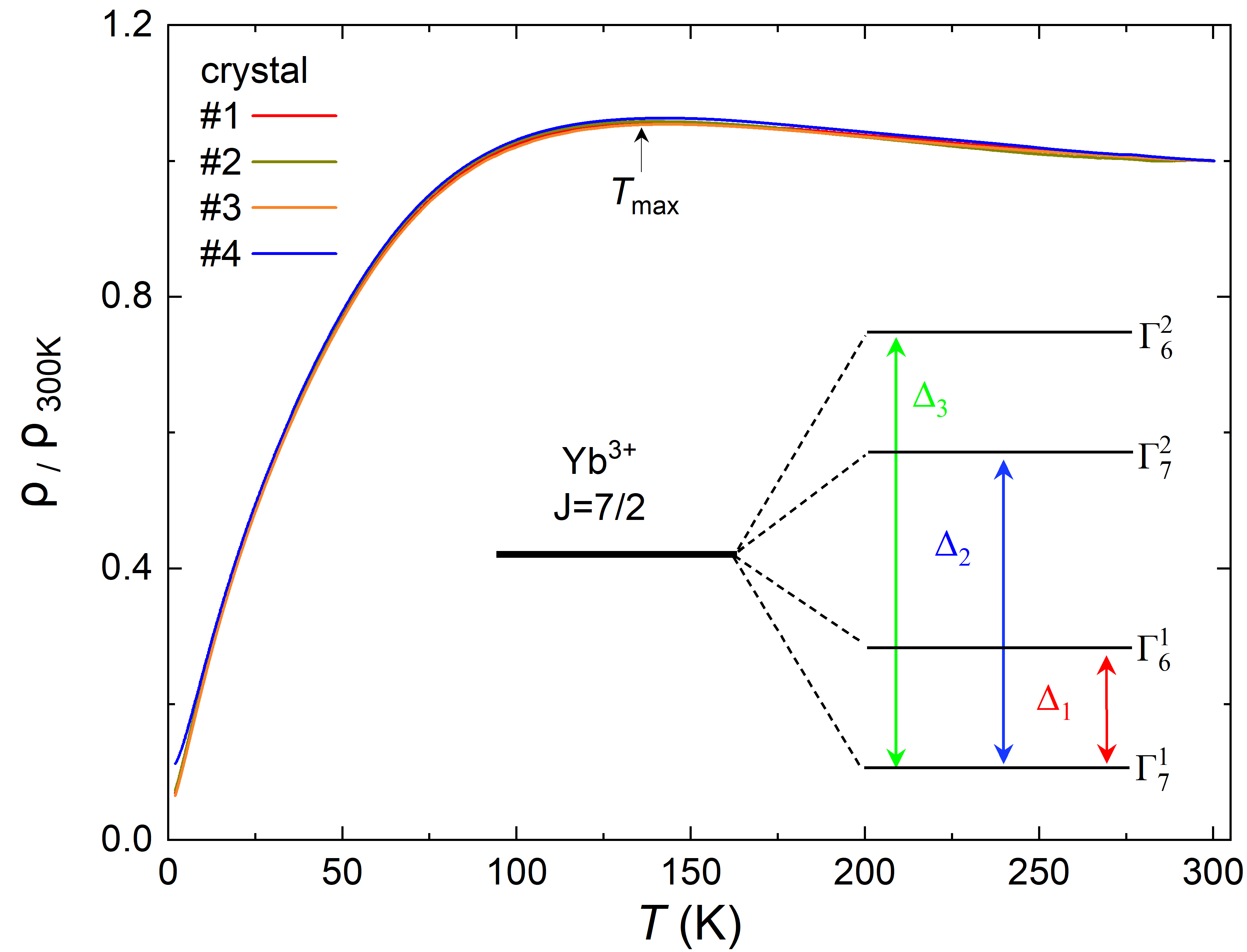}
	\caption{Normalized electrical resistance of four different single crystals of YbRh$_{2}$Si$_{2}$ at ambient conditions (no strain). The arrow indicates the position of the maximum temperature $T_{\rm max}$. The inset depicts schematically the splitting of the $Yb^{\rm 3+}$ $J=7/2$ multiplett in the tetragonal crystal electric field into four Kramers doublets~\cite{Inelastic}.}
	\label{figure3}
\end{figure}

Single crystals of YRS were grown using the In-flux technique~\cite{int3,Krellner2012}. Yb ingots (Ames, 99.99\%), Rh powder (Heraeus, 99.95\%, Si pieces (Alfa Aesar, 99.99\% and In shots (Alfa Aesar, 99.9999\%) were put together in a cylindrical Al$_{2}$O$_{3}$ crucible with 96\% mol In and 4\% mol YbRh$_{2}$Si$_{2}$. The Yb was handled and stored in an argon filled glovebox (O$_2< 1$ ppm, H$_2$O $<$ 1 ppm). The Al$_{2}$O$_{3}$ crucible was enclosed inside a Ta tube and sealed using arc welding under 0.5 bar Ar~\cite{Canfield}. The growth was carried out in an Ar-filled, vertical tube furnace with the Ta container wrapped in Zr foil in order to prevent oxidation.  The elements were heated up to $1480\degree$C followed by slow cooling to $T=950\degree$C over 5 days. After the growth, the plate-like single crystals were extracted by dissolving the In-rich flux in diluted hydrochloric acid solution. The orientation was determined using Laue back reflection. The plates were sliced to bars along the [100] (a-axis) direction with a length between 1.7 to 2~mm and  width in between 100-250~$\mu$m, using an DIDRAS diamond wire saw cutter and polished to thickness 40-80~$\mu$m. Electrical contacts with Au wires of thickness 25~$\mu$m were made using two component H20 silver epoxy from EPOTEK, cured at 100$\degree$C. The four-probe electrical resistance measurements as function of temperature and strain were performed in a PPMS (Quantum Design) with the strain cell (FC100 from Razorbill) attached to the modified P450 probe (Quantum Design). Thermal anchoring of the cell is achieved by silver foil and wires. As depicted in Fig.~\ref{figure2}, the rectangular bar shaped crystals were mounted onto the clamps of the stress cell using Stycast 2850FT two component epoxy such that strain on the crystal was along the [100] direction. The horizontal force $F$ applied to the crystal for a given applied voltage on the piezo-stacks was determined by measuring the change in capacitance of the inbuilt pre-calibrated capacitive force sensor on the FC 100 cell~\cite{razorbill}. The uniaxial pressure follows from $P=F/A$ with the sample cross section $A$. Within the elastic regime and assuming a constant Young's modulus $Y=189$~GPa for YRS~\cite{YRSyoungmodulas} the strain is given by $\epsilon=P/Y$. To minimize strain inhomogeneity between the voltage contacts, it is crucial to mount samples with high length-to-width aspect ratios and place the contacts sufficiently far from the edges, since the strain inhomogeneity decays exponentially towards the center. Comparison with model calculations~\cite{Hicks2014Rsi} indicates an inhomogeneity below 5\%. The piezo-voltage was applied after stabilizing 160~K. Then the sample resistance was detected upon cooling down to 2 K and subsequent warming back to 160 K with a rate of $1$~K/min. No thermal hysteresis within the error bar of $\pm 1$~K was detected. The design of the cell compensates the thermal contraction of the setup and ensures a constant force. Three crystals were studied under tensile and one more under compressive strain.

\section{Results and Discussion} 

\begin{figure*}
	\includegraphics[scale=0.4]{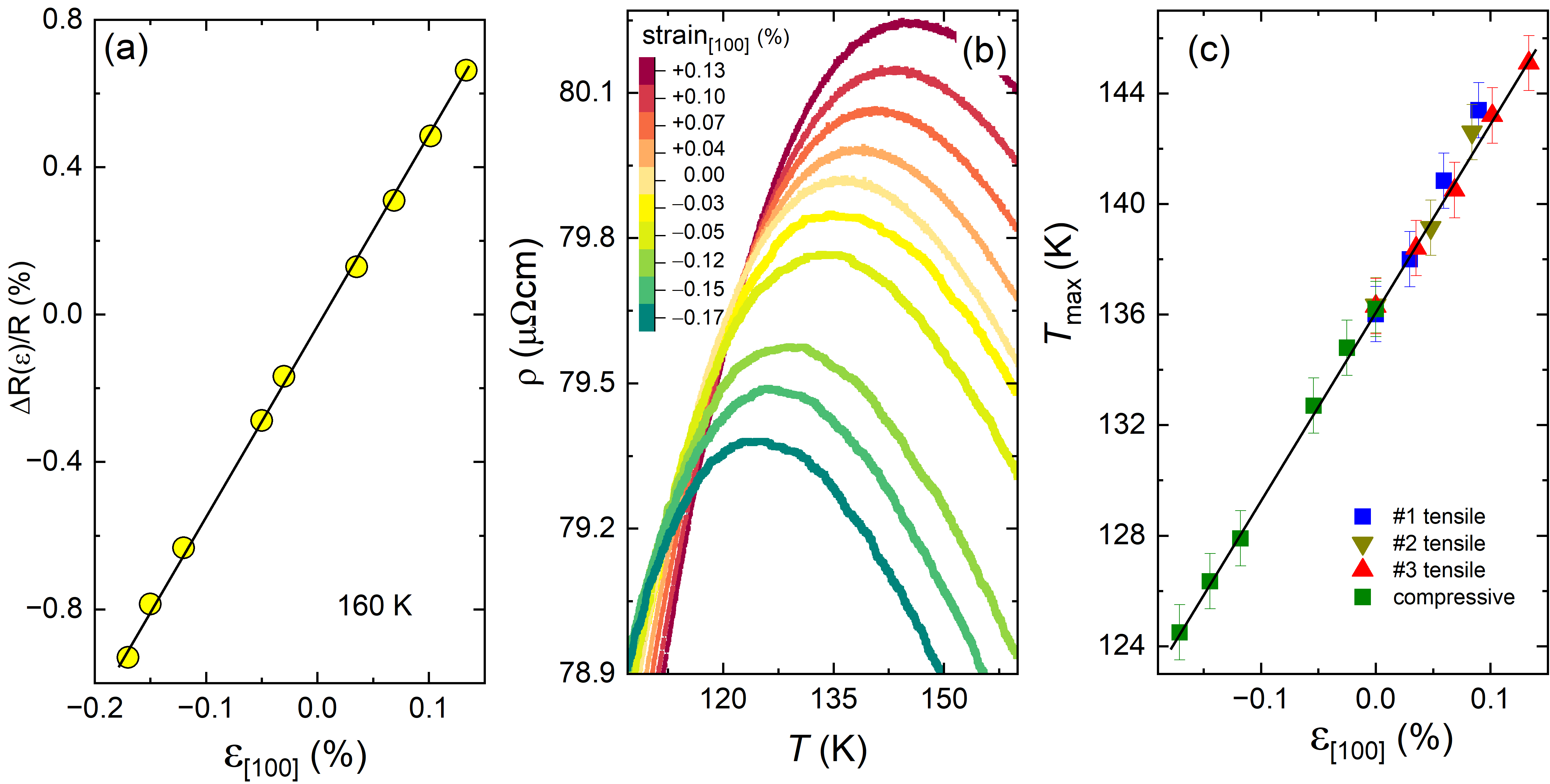}
	\caption{(a): Linear evolution of the elastoresistance of YbRh$_{2}$Si$_{2}$ at 160~K as a function of tensile (crystal 3) and compressive strain along the crystallographic [100] direction. (b):  Effect of positive (tensile) and negative (compressive) strain (along [100]) on the resistance maximum of YbRh$_{2}$Si$_{2}$ crystal $\#3$. (c): Shift of $T_{\rm max}$ with strain (along [100]) for different YbRh$_{2}$Si$_{2}$ crystals. Error bars indicate thermometry accuracy of 1~K.}
	\label{figure4}
\end{figure*}

\begin{figure}
 	\includegraphics[scale=0.28]{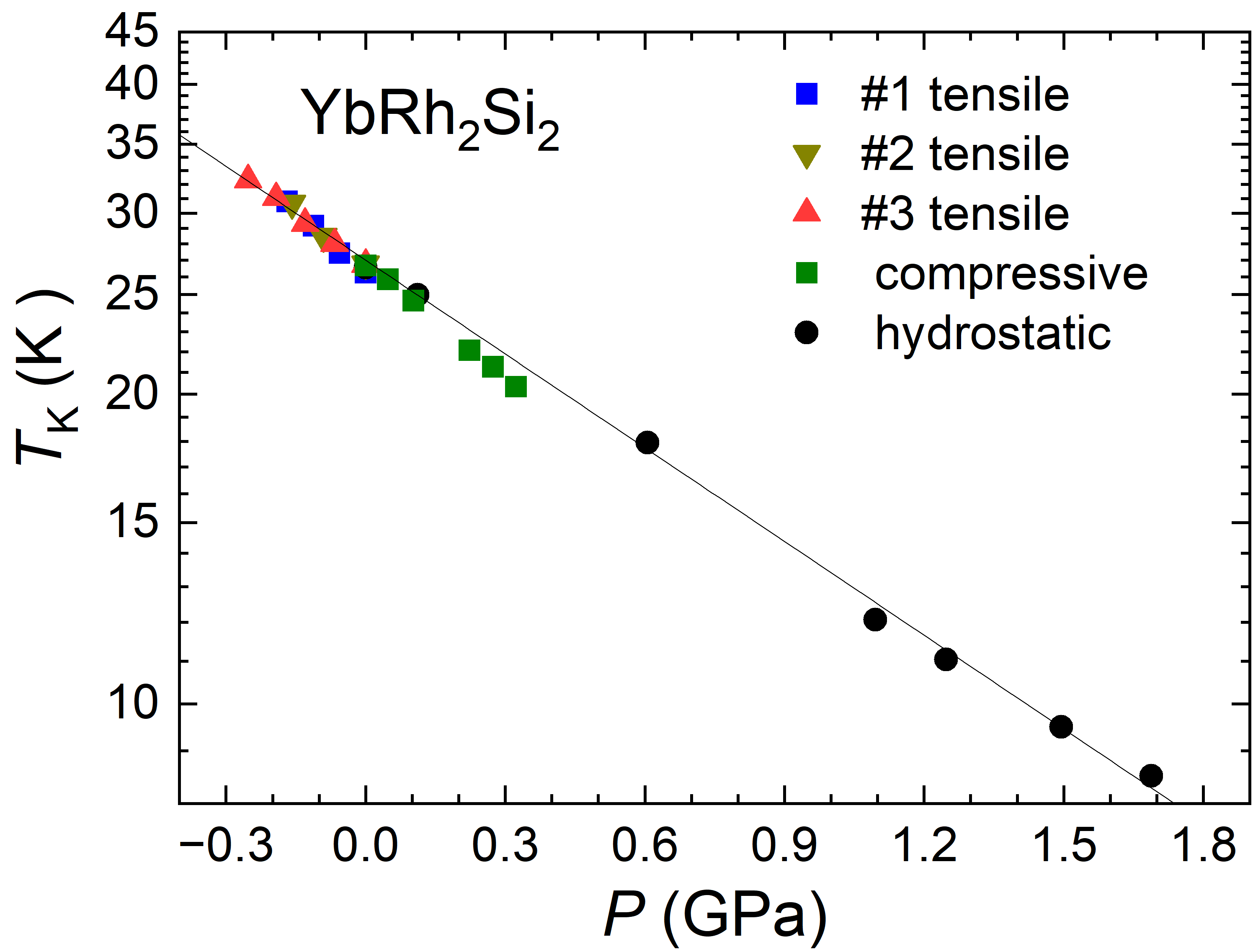}
 	\caption{Hydrostatic pressure dependence of Kondo temperature $T_{\rm K}$ of YbRh$_{2}$Si$_{2}$ from Tokiwa {\it et al.}~\cite{PRLTokiwa} (solid black circles) compared to $T_{\rm K}$ under tensile and compressive strain along the [100] direction, as calculated using equation~(\ref{eq1}) with constant adjustment factor to match with $T_{\rm K}=27$~K at $P=0$.}
 	\label{figure5}
 \end{figure}

Fig.~\ref{figure3} shows the temperature dependence of the electrical resistance of four different single crystals at ambient strain down to 1.8~K. After passing a characteristic maximum around $T_{\rm max}=136$~K the resistance decreases upon cooling and reaches $\rho_{300{\rm K}}/\rho _{1.8{\rm K}}$ values in between 8 to 14 similar as reported earlier~\cite{int4}. The resistivity maximum is characteristic for heavy-fermion metals and arises from incoherent Kondo scattering on the ground state and the excited crystal electric field (CEF) levels (cf. the inset of Fig.~\ref{figure3}) of the f-electrons~\cite{Cornut1972}. Note, that due to the CEF effect the resistivity maximum can appear far above the Kondo temperature of the ground state doublet~\cite{Lassailly1985}. Inelastic neutron-scattering experiments on YRS powder have found CEF excitations of the Yb$^{3+}$ ions at approximately 17, 25 and 43~meV~\cite{Inelastic}. Since the highest level corresponds to temperatures well above room temperature, we consider the Kondo effect on the ground state and first two excited doublets (with excitations energies $\Delta_1$ and $\Delta_2$ in Kelvin, respectively), yielding according to Hanzawa et al.~\cite{Hanzawa1985}, a "high-Kondo temperature"
 \begin{equation}
 	\begin{aligned}
 		T_{\rm K}^{\rm h}= \sqrt[3]{T_{\rm K}\Delta_1\Delta_2},
 	\end{aligned}
 \label{eq1}
  \end{equation}
where $T_{\rm K}$ denotes the Kondo temperature for the ground state doublet, which is around 27~K for YRS~\cite{PRLTokiwa}. Using this value and splittings of $\Delta_1=197$~K and $\Delta_2=290$~K yields $T_{\rm K}^{\rm h}=112$~K, which is about 20\% lower than the maximum temperature $T_{\rm max}$ of the electrical resistance at ambient strain. However, it should be noted that the exact position of $T_{\rm max}$ is also influenced by the unknown phonon background to $\rho(T)$ and a previous attempt to subtract such background by the analysis of $\rho(T)$ data of nonmagnetic LuRh$_2$Si$_2$ indeed revealed a lower value of $T_{\rm max}$, close to the pronounced negative peak in the thermoelectric power at 80-100~K in Lu$_{1-x}$Yb$_{x}$Rh$_2$Si$_2$~\cite{Koehler2008}. Since we follow the {\it relative} change of energy scales in YRS with strain, we prefer to directly analyze our raw electrical resistance data without any background subtraction and will later use a scaling factor to adjust the strain dependence of the Kondo temperature at zero strain to $T_{\rm K}=27$~K. Note that the phonon contribution to the electrical resistance scales with the Debye temperature, whose change under $10^{-3}$ strain is tiny. Additionally, tetragonal YbRh$_2$Si$_2$ is far from any structural phase transitions. Thus, strain induced changes of $\rho(T)$ result mostly from the 4f contribution and not primarily from phonons.

Next, we discuss the change of the electrical resistance of YbRh$_{2}$Si$_{2}$ with tensile and compressive strain along the [100] direction. As shown in Fig.~\ref{figure4}(a) the electrical resistance of YbRh$_{2}$Si$_{2}$ at 160 K shows a weakly linear dependence on uniaxial tensile and compressive strain, proving a full transmission of the applied strain to the crystal. It is worth to mention that crystals are prone to crack formation under tensile strain, which currently prevents us from reaching tensile strain values exceeding 0.13\%. The temperature dependence of the resistance between 105 and 160 K at these applied strains, in addition to the zero-strain curve, is displayed in Fig.~\ref{figure4}(b). $T_{\rm max}$ shows a pronounced shift towards higher values with increasing tensile strain, from 136 to 145~K for $\epsilon=0.13\%$. A very similar shift is confirmed on two other crystals, as depicted in Fig.\ref{figure4}(c). Note, that the linear strain dependence of  $T_{\rm max}$ only holds as long as the crystals are not yet cracked.
Under compressive strain, $T_{\rm max}$ shifts towards low temperature from 136 to 124.5~K  for $\epsilon=0.17\%$ and follows the same linear strain dependence in continuation with tensile strain as depicted in Fig.\ref{figure4}(c).

From the above general consideration of Yb-based Kondo metals, such rapid increase of $T_{\rm max}$ towards high temperature under tensile strain suggests an expected enhancement of the Kondo interaction. It is also qualitatively consistent with the suppression of $T_{\rm max}$ and $T_{\rm K}$ in YbRh$_{2}$Si$_{2}$ with hydrostatic pressure~\cite{DIONICIO2005,PRLTokiwa}, as discussed below.


Although the observed tensile and compressive strain dependences suggest that the shift of $T_{\rm max}$ reflects a change of the Kondo temperature, we also need to consider the strain effect on the CEF splitting, since as discussed above, the latter influences the position of the resistance maximum.
To estimate the change of the CEF splitting with strain, the elastic deformation of the crystal lattice is considered, following Refs.~\cite{gati2023elastocaloric,Kuromaru}. The strain along the tetragonal [100] direction leads to an orthorhombic distortion. Hence, we evaluate the following Hamiltonian:
\begin{equation}
\begin{aligned}
H_{\rm total} =H ^{\rm tetra}_{\rm CEF} + H^{\rm ortho}_{\rm ME}
\end{aligned} 
\label{eq3}
\end{equation}
Here,
\begin{equation}
\begin{aligned}
H^{\rm tetra}_{\rm CEF} =\alpha(B ^{0}_{2}O ^{0}_{2})+\beta(B ^{0}_{4}O ^{0}_{4}+B ^{4}_{4}O ^{4}_{4})\\
+\gamma(B ^{0}_{6}O ^{0}_{6}+B ^{4}_{6}O ^{4}_{6})
\end{aligned} 
\end{equation} 
{where $O^m_n$ are Stevens operator equivalents, $B^m_n$ are the CEF parameters taken from Kutuzov {\it et al.}~\cite{Kutuzov_2011}, while Stevens' multiplicative factors for Yb ions are $\alpha$ = 2/63, $\beta$ = 2/1155, $\gamma$ = 4/27027. Assuming the magnetoelastic effect $H^{\rm ortho}_{\rm ME}$ is small, we only consider first-order perturbation (i.e., to the first order of strain-CEF coupling) as a starting point. We obtain the ground state doublet $|\Psi_\pm \rangle$, and the first excited doublet $|\Psi'_\pm \rangle$ by diagonalizing eq. (3.3) which couples $|J, J_{\rm z}\rangle$ as follows:
\begin{equation}
|\Psi_\pm \rangle= a|7/2, \pm 5/2\rangle + b|7/2, \mp 3/2\rangle
\end{equation}
\begin{equation}
|\Psi'_\pm \rangle= a'|7/2, \pm 7/2\rangle + b'|7/2, \mp 1/2\rangle
\end{equation}
Normalization requires $a^2+b^2=1$ and $a'^2+b'^2=1$. Orthorhombic distortion includes $O^{2}_{2}, O^{2}_{4}, O^{2}_{6}, O^{6}_{6}$ in addition to the five other terms that are also included in $H^{\rm tetra}_{\rm CEF}$. Since the operator $O^2_n$ ($O^6_n$) only couples to the state of $\Delta J_{\rm z}=\pm 2$ ($\pm 6$), the matrix elements $\langle \Psi |O^2_n (O^6_n)|\Psi \rangle$ vanish exactly.  In other words, a small orthorhombic distortion does not add more terms to the original CEF environment $H^{\rm tetra}_{\rm CEF}$ because of the incompatible symmetry between the CEF doublets and $O^2_n$, $O^6_n$ terms, as mentioned above. Therefore, the CEF wavefunctions retain their structure. Incidentally, the effects from $O^2_n$ and $O^6_n$ are no longer negligible for second-order perturbation because the off-diagonal matrix elements between the CEF ground state and the first excited state $\langle \Psi |O^2_n (O^6_n)|\Psi' \rangle$ have finite values. For first-order perturbation we could write eq (\ref{eq3}) as follows:
\begin{equation}
\begin{aligned}
H_{\rm total} =H ^{\rm tetra}_{\rm CEF}+g_{xx} \epsilon_{xx}(O^{0}_{2}+\Delta g_{04}O^{0}_{4}+\Delta g_{44}O^{4}_{4}\\ +\Delta g_{06}O^{0}_{6}+  \Delta g_{46}O^{4}_{6})
\end{aligned} 
\end{equation}
where $g_{xx}$ is the magnetoelastic coupling constant, $\epsilon_{xx}$ is the applied strain along the [100] direction, and $\Delta\textit{g}_{mn}$ is the weight on each Stevens operator equivalents. We further simplify the equation by dropping the higher order terms $O^{m}_{4}$ and $O^{m}_{6}$, since they should contribute much less than the quadrupolar term. For example, in the case of the rare-earth intermetallic compound TmSb, the magnetoelastic coupling with  $O_{4}^{m}$ is only 16~mK, which is three orders of magnitude smaller than the coupling with the quadrupolar term which is 20 K~\cite{Luthi2005}. 

Diagnolization of the simplified Hamiltonian
\begin{align}
  H ^{\rm tetra}_{\rm CEF}+g_{xx}\epsilon_{xx} O^{0}_{2}
\end{align}
allows then to calculate the approximate strain effect on the energy shifts between the protected Kramers doublets. Upon first-order perturbation, the energy for the ground state doublet is
\begin{equation}
E = E_0 + g_{xx}\epsilon_{xx} \langle \Psi |O^0_2 |\Psi \rangle
\end{equation}
and for the first excited doublet
\begin{equation}
E' = E'_{0}+g_{xx}\epsilon _{xx} \langle \Psi '|O^{0}_{2}|\Psi ' \rangle
\end{equation}
We define the energy gap between the ground state and the first excited state as $\Delta_1 = E'_0-E_0$ where $E_0$ ($E'_0$) is the unperturbed energy of the ground state doublet (first excited doublet), and the energy gap after introducing the first-order perturbation as $\Delta_1^{(1)} = E'-E$. The energy gap between the first and the second excited doublet with (without) perturbation $\Delta_2^{(1)}$ ($\Delta_2$) can be defined in a similar manner.

Magnetoelastic coupling constants can be determined by measurements of the ultrasound velocity. For YRS there exist to the best of our knowledge no such data yet, likely because available thin crystal plates are not suitable for such experiments. As a reference, we discuss values of other Yb-based Kondo lattices. For YbPtBi, it is estimated to be around 10 K~\cite{gati2023elastocaloric}, while for YbCo$_2$Zn$_{20}$ and YbRh$_2$Zn$_{20}$, the fitting of ultrasound data yielded 1 K \cite{Nakanishi_2009} and 131 K \cite{Nakanishi_Rh_2009} for the largest modes, respectively. Obviously, the magnetoelastic coupling can vary substantially even for materials with the same crystal structure and similar magnetic ions. We thus used a value of 131 K to minimize the risk of underestimating magnetoelastic coupling to the best of our ability. The result of the model calculation indicates that both $\Delta_1 ^{(1)}$ and $\Delta_2 ^{(1)}$ are only shifted around $1 \%$ with respect to the unperturbed $\Delta _1$ and $\Delta _2$, respectively, for  strain $\epsilon_{xx}$ up to $\pm 0.2\%$. The small shift of the CEF energy justifies our approximation based on first-order perturbation. For YRS, such a minute variation of the excitation energies cannot explain the significant shift of the maximum temperature of the electrical resistance, which thus can indeed be related to the strain-induced change of the Kondo temperature.

Using equation~(\ref{eq1}), $T_{\rm K}$ can be calculated from the experimentally determined values of $T_{\rm max}$ and the uniaxial pressure dependence of $\Delta_1$ and $\Delta_2$ from the above calculation.  Calculated values of $T_{\rm K}$ with a constant prefactor adjusted to match 27~K at ambient conditions~\cite{PRLTokiwa} is shown in Fig.~\ref{figure5}. Tensile strain, corresponding to negative hydrostatic pressure, clearly enhances the Kondo temperature $T_{\rm K}$ up to 32~K. According to the previous study of $T_{\rm N}$ under hydrostatic pressure, the extrapolated AFM QCP in YRS is located around $-0.3$~GPa~\cite{Mederle2002}. Furthermore, the extrapolation of the hydrostatic pressure dependence of $T_{\rm K}$ \cite{PRLTokiwa} to $-0.3$~GPa yields a $T_K$ of 33~K, which is very close to the maximal value we have obtained under tensile strain. This suggests that the nature of the AFM QCP in pure YRS {\it at zero field} can be studied in future tensile strain experiments. It requires to overcome demanding challenges of reaching milli-Kelvin temperatures due to the large mass and poor thermal conductance of the piezo-strain apparatus. A study of the temperature dependence of the electrical resistance, yielding the phase boundary $T_{\rm N}(\epsilon)$, the crossover to Landau Fermi liquid behavior $T_{\rm FL}(\epsilon)$, as well as the strain dependence of the quasiparticle scattering cross section is very interesting, since distinct behavior compared to the case of field-tuning, which polarizes the AFM ordered moments could be expected, as discussed in the introduction. The effect of {\it compressive} uniaxial as well as biaxial strain, enhancing $T_{\rm N}$, has been successfully determined earlier by electrical resistance measurements on thin microstructured YRS meanders, glued with epoxy to a silver holder~\cite{Stepke}.

In summary, we have successfully realized tensile piezo-strain tuning of the prototypical heavy fermion system YbRh$_{2}$Si$_{2}$. The characteristic maximum in the temperature dependence of the electrical resistance has been significantly enhanced for  0.13\% tensile strain. Our CEF calculations reveal that the change of $T_{\rm max}$ is primarily caused by the change of the Kondo temperature. The significant enhancement of $T_{\rm K}$ with tensile strain allows to tune clean undoped YRS across its zero-field QCP in future milli-Kelvin experiments. While compressive piezo-strain has recently been used for enhancing the Kondo interaction in a Ce-based heavy-fermion metal~\cite{Cheng} our work provides the path to explore hidden regimes in heavy-fermions by tensile strain.

\section*{ACKNOWLEDGMENT}
We are grateful to Elena Gati, Burkhard Schmidt, Hiroaki Kusunose, and Tatsuya Yanagisawa for valuable discussions regarding the strain dependence of the CEF splitting in Kondo systems. S.N. Panja is supported by the Alexander von Humboldt Foundation.
\bibliography{Bibliography}
\end{document}